\documentstyle{pss}
\input epsf.tex
\begin{document}
\title{Spin waves in La$_2$CuO$_4$: band structure and correlation effects}

\author{{\sc N. M. R. Peres}\footnote{( Corresponding author;
e-mail: peres@fisica.uminho.pt)} (a,c),  and {\sc M. A. N. Ara\'ujo} (b,c)}

\address{
(a) Departamento de F\'{\i}sica da Universidade do Minho, Campus Gualtar, 
P-4700-320 Braga, Portugal.
\\\\
(b) Departamento de F\'{\i}sica, Universidade de \'Evora,
Rua Rom\~ao Ramalho, 59, P-7000-671 \'Evora, Portugal. 
\\\\
(c) Centro de F\'{\i}sica da Universidade do Minho, Campus Gualtar, 
P-4700-320 Braga, Portugal.}

\submitted{March 22, 2002} \maketitle

\hspace{9mm} Subject classification: 75.30.Ds, 71.10.Fd, 75.40.Gb

\begin{abstract}
We calculate the antiferromagnetic spin wave dispersion in the 
half-filled (electronic density $n=1$) 
Hubbard model for a two-dimensional square lattice,
using the 
random phase approximation (RPA)
in a broken symmetry (spin density
wave) ground state. 
Our results for the
spin wave dispersion, $\omega(\vec q)$,
are compared with high-resolution inelastic 
neutron scattering performed on La$_2$CuO$_4$. The effects of different band
structures and different values of the on-site Coulomb interaction 
on the spin wave spectrum is studied. Particular attention is put
on the high energy dispersion values
$\omega(\pi/2,\pi/2)$ and $\omega(0,\pi)$.  
\end{abstract}

\section{Introduction:}

In two recent  papers, \cite{coldea,ronnow}
high-resolution inelastic neutron scattering measurements 
have been performed
on two different two-dimensional spin 1/2 quantum antiferromagnets. These
are 
copper deuteroformate tetradeuterate (CFTD) and La$_2$CuO$_4$. Surprisingly,
the dispersion at the zone boundary that has been 
observed in the two materials,
does not agree with spin-wave theory predictions \cite{igarashi}. 
Moreover the 
amount of dispersion is not the same for both materials. In CFTD
the dispersion is about 6\% from $\omega(\pi/2,\pi/2)$ to 
$\omega(\pi,0)$, whereas in  La$_2$CuO$_4$ it is about -13\% along
the same direction. In the case of CFTD the dispersion at the zone boundary
can be explained using the nearest-neighbor Heisenberg model alone,
\cite{ronnow} and high precision quantum Monte Carlo simulations
have confirmed  that it is so.\cite{sandvik} On the other hand, an
explanation for the observed dispersion in  La$_2$CuO$_4$  has been proposed 
\cite{coldea}  using   an extended Heisenberg model  
\cite{takahashi,macdonald} involving first-,
second-, and third-nearest-neighbor interactions
as well as interactions among four spins. 

In a previous paper \cite{nmr}, we have shown that it possible
to obtain the observed dispersion difference of -13\% for 
La$_2$CuO$_4$  using the single band Hubbard model at half filling,
with nearest neighbor hopping.
In our formulation the extended Heisenberg model
used in ref. \cite{coldea}
is incorporated by means of  
the virtual excursions of the electrons on the lattice.
Fitting our results  to the experimental data the 
obtained values of $U$ and $t$ agree well with those
of ref. \cite{coldea} and where confirmed by Quantum Monte Carlo
calculations in the Hubbard model \cite{pinaki}.  

In this paper we generalize our previous study incorporating in 
the calculations the effect of a second nearest neighbor hopping
$t'$ in the electronic spectrum (in high-$T_c$ materials the 
ratio $\vert t'/t \vert $ ranges roughly from 0.1 to 0.5).
The effect of $U$ on the spin wave dispersion at the special
points  $\omega(\pi/2,\pi/2)$ and $\omega(0,\pi)$
is also studied.

\section{Model Hamiltonian:}

The Hubbard model for a square lattice of $N$ sites is defined as
\begin{eqnarray}
H&=&\sum_{\vec k,\sigma}[\epsilon(\vec k)-\mu]
c^{\dag}_{\vec k,\sigma}c_{\vec k,\sigma}+
U\sum_i
c^{\dag}_{i,\uparrow}c_{i,\uparrow}c^{\dag}_{i,\downarrow}c_{i,\downarrow}\,,
\label{hubbard}
\end{eqnarray}
where $\epsilon(\vec k)$ defines the 
energy dispersion for independent
electrons. In this
work we consider two different electronic energy dispersions 
given by 
\begin{equation}
\epsilon(\vec k)=-2t\cos k_x-2t\cos k_y 
\label{d1}
\end{equation}
and by
\begin{equation}
\epsilon(\vec k)=-2t\cos k_x-2t\cos k_y - 4t'cos(k_x)cos(k_y).
\label{d2}
\end{equation}
The first energy dispersion has  
the nesting vector $\vec Q=(\pi,\pi)$; the second one is not nested. 

The broken symmetry state is introduced by considering
the existence of an off-diagonal Green's function given by 
\begin{equation}
F_{\sigma}(\vec p;\tau-\tau')=-< T_{\tau}
c_{\vec p\pm\vec Q,\sigma}(\tau)c^\dag_{\vec p,\sigma}(\tau')>\,.
\end{equation}
in addition to the usual   Green's function: 
\begin{equation}
G_{\sigma}(\vec p;\tau-\tau')=-< T_{\tau}
c_{\vec p,\sigma}(\tau)c^{\dag}_{\vec p,\sigma}(\tau')>\,.
\end{equation}
At the mean field level the Fourier transform
of these two  Green's functions are given by

\begin{eqnarray}
G(\vec p,i\omega_n)&=&\frac {u_{\vec p}}{i\omega_n-E_+(\vec p)}+
\frac {v_{\vec p}}{i\omega_n-E_-(\vec p)}\,,\\
F_{\sigma}(\vec p,i\omega_n)&=&\frac {\tilde u_{\vec p,\sigma}}{i\omega_n-E_+(\vec p)}+
\frac {\tilde v_{\vec p,\sigma}}{i\omega_n-E_-(\vec p)}\,,
\end{eqnarray}
where the energies $E_{\pm}$ are given by
\begin{equation}
E_\pm(\vec p)=\frac {\xi(\vec p)+\xi(\vec p+\vec Q)}2 +U\frac n2 \pm
\frac 1 2\sqrt{[\xi(\vec p)-\xi(\vec p+\vec Q)]^2+U^2m^2}\,,
\end{equation}
$\xi(\vec p)=\epsilon(\vec p)-\mu$, and the coherence factors read
\begin{equation}
u_{\vec p}=\frac {E_+-\xi(\vec p+\vec Q)-Un/2}{E_+-E_-}\,,
\hspace{0.5cm}
v_{\vec p}=\frac {E_+-\xi(\vec p)-Un/2}{E_+-E_-}\,,
\end{equation}
and
\begin{equation}
\tilde u_{\vec p,\sigma}=-\frac {Um\sigma/2}{E_+-E_-}\,,
\hspace{0.5cm}
\tilde v_{\vec p,\sigma}=\frac {Um\sigma/2}{E_+-E_-}\,.
\end{equation}

\section{Spin susceptibility and spin waves:}

In order to describe the spin dynamics of the system we consider 
the  transverse spin susceptibility
$ \chi_{-+}(\vec q,i\omega_n)$,  which is defined
as  
\begin{equation}
\chi_{-+}(\vec q,i\omega_n)=\mu^2_B\int_0^\beta d\,\tau e^{i\omega_n\tau}
< T_\tau S^-(\vec q,\tau)S^+(\vec q,0)>\,,
\end{equation}
where $\beta=1/T$  is the inverse temperature,
$T_\tau$ is the chronological order operator (in imaginary time),
$S^-(\vec q)= \sum_{\vec p}c^\dag_{\vec p,\downarrow}
c_{\vec p+\vec q,\uparrow}$
and $S^+(\vec q)=[S^-(\vec q)]^\dag$.
The above expression can be written as
\begin{eqnarray}
&&
\chi_{+-}(\vec q,i\omega_n)=\mu^2_B
\sum_{n=0}^\infty
\int_0^\beta d\,\tau\sum_{\vec p,\vec p'}e^{i\omega_n\tau}
\nonumber\\
&&< T_\tau [-\int_0^\beta d\,\bar\tau H_U(\bar\tau)]^n
c^\dag_{\vec p,\downarrow}(\tau)c_{\vec p+\vec q,\uparrow}(\tau)
c^\dag_{\vec p'+\vec q,\uparrow}(0)c_{\vec p',\downarrow}(0) >_{d.c.}\,,
\label{chi}
\end{eqnarray}
where $d.c.$ stands for differently connected diagrams.
The susceptibility is evaluated at the RPA level \cite{nmr} and the spin
wave spectrum $\omega(\vec q)$ is determined from the poles of 
$\chi_{+-}(\vec q,\omega+i0^+)$.
\begin{figure}[h]
\begin{tindent}
\epsfxsize=7cm
\epsffile{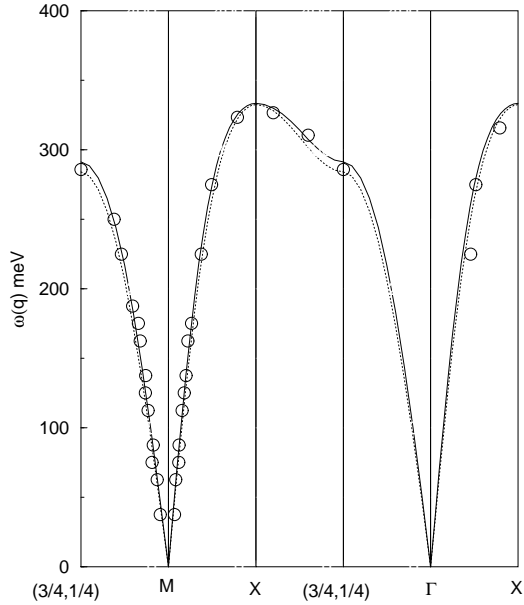}
\end{tindent}
\caption{Spin wave dispersion, in meV, along 
high symmetry directions in the Brillouin zone. 
The circles are the data reported in
Ref. \cite{coldea} at $10$ K. 
The {\bf dashed line} is the analytical result (at 0 K) for 
$U=1.8$ eV, $t=0.295$ eV. 
The {\bf solid line} is the analytical result (at 0 K)
$U=2.1$ eV, $t=0.34$ eV, and $t'/t=-0.25$.
The momentum is in units of $2\pi$ and
$M=(1/2,1/2)$, $X=(1/2,0)$, and $\Gamma=(0,0)$.}
\label{fig1}
\end{figure}

In Figure~\ref{fig1} the spin wave spectrum is plotted along high
symmetry directions of the Brillouin zone for both dispersions
(\ref{d1}) and (\ref{d2}). For the case of the dispersion (\ref{d1})
the values of $t$ and $U$ giving the best fit to the experimental
data are, for $T=0$ K,  $U=1.8$ eV, $t=0.295$ eV,
with $U/t=6.1$ (we note that there is no measurable change
between the calculation at  $T=0$ K and $T=10$ K). 
For the dispersion
(\ref{d2}), which does not present nesting, the experimental data
can be fitted using $U=2.1$ eV, $t=0.34$ eV, with
$U/t=6.2$, and $t'/t=-0.25$. These last set of values
agree with those determined in ref. \cite{avinash}, from 
an perturbative calculation of the poles of $\chi_{+-}(\vec q,\omega+i0^+)$.  
It is clear from these results that the introduction of a more
realistic band structure, given by (\ref{d2}), leads to larger
values of $t$ and $U$.  

\begin{figure}[h]
\begin{tindent}
\epsfxsize=7cm
\epsffile{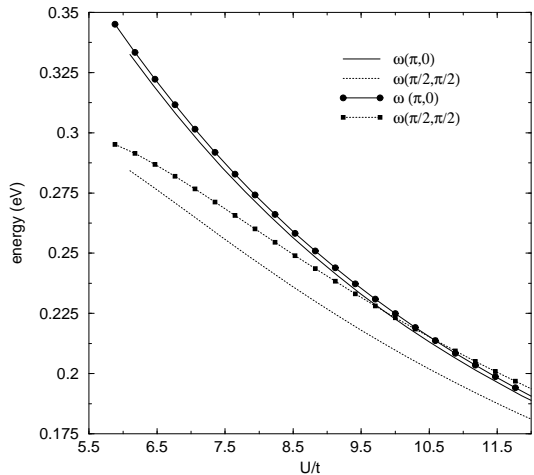}
\end{tindent}
\caption{Variation of $\omega(0,\pi)$ and $\omega(\pi/2,\pi/2)$
as function of $U/t$. The lines with circles and squares refer
to the case where $t'/t=-0.25$; the other lines refer to the case
$t'=0$.}
\label{fig2}
\end{figure}

Let us now consider the effect of the Coulomb interaction $U$
on the energy difference $\omega(0,\pi)-\omega(\pi/2,\pi/2)$.
Experimentally 
the dispersion, in CFTD, is about 6\% from $\omega(\pi/2,\pi/2)$ to 
$\omega(\pi,0)$, whereas in  La$_2$CuO$_4$ it is about -13\% along
the same direction. The question is whether it is possible to
obtain these very different behaviors from a single model Hamiltonian.
In Figure~\ref{fig2} we study the evolution of 
$\omega(0,\pi)$  and $\omega(\pi/2,\pi/2)$ as function of $U$
for the dispersions (\ref{d1}) and (\ref{d2}). For the dispersion
(\ref{d1}) it is clear that $\omega(0,\pi)$  is always
larger then $\omega(\pi/2,\pi/2)$, 
except in the limit $U\rightarrow\infty$ where they become equal.
This behavior was confirmed by Quantum Monte Carlo \cite{pinaki}. 
On the other hand, for the dispersion (\ref{d2}),
$\omega(0,\pi)$  and $\omega(\pi/2,\pi/2)$ become equal for a finite
value of $U/t$, which for the chosen parameters is 
$U/t\simeq 10.5$. For larger values of $U/t$ the dispersion at 
$\omega(\pi/2,\pi/2)$ becomes larger than $\omega(0,\pi)$.

In conclusion, our results shown that it is possible to fit the
spin wave spectrum with the $t-t'-U$ Hubbard model and that the
two different behaviors for $\omega(0,\pi)$ and 
$\omega(\pi/2,\pi/2)$ observed in La$_2$CuO$_4$ and in 
Cu(DCOO)$_2$.4D$_2$O follow from a single model Hamiltonian.


\begin{references}
\bibitem{coldea} 
R. Coldea, S. M. Hayden, G. Aeppli, T. G. Perring, C. D. Frost, T. E.
Mason, S.-W. Cheong, and Z. Fisk, Phys. Rev. Lett. {\bf 86}, 5377 (2001).
\bibitem{ronnow}
H. M. R{\o}nnow, D. F. McMorrow, R. Coldea, A. Harrison, I. D. Youngson, 
T. G. Perring, G. Aeppli, O. Sylju{\aa}sen, K. Lefmann, and C. Rischel,
Phys. Rev. Lett. {\bf 87}, 37202 (2001).
\bibitem{igarashi} J. Igarashi, Phys. Rev. B {\bf 46}, 10763 (1992).
\bibitem{sandvik} Anders Sandvik and Rajiv R. P. Sing, Phys. Rev. Lett.
{\bf 86}, 528 (2001).
\bibitem{takahashi}
M. Takahashi, J. Phys. C {\bf 10}, 1289 (1977).
\bibitem{macdonald}
A. H. MacDonald, S. M. Girvin, and D. Yoshioka, Phys. Rev. B {\bf 41},
2565 (1990); {\bf 37}, 9753 (1988).
\bibitem{nmr}
N. M. R. Peres and M. A. N. Ara\'ujo, Phys. Rev B {\bf 65},
132404 (2002).
\bibitem{pinaki} Pinaki Sengupta, Richard T. Scalettar, and
Rajiv R. P. Singh, Phys. Rev. B {\bf 65}, 132404 (2002).
\bibitem{avinash}
Avinash Singh annd Pallab Goswami, Phys. Rev. B {\bf 66}, 092402 (2002).
\end{references}
\end{document}